\documentclass{article}    

\usepackage{graphicx}

\title{\textbf{A Foundation Theory of Quantum Mechanics}}  
\author{Richard Mould\footnote{Department of Physics and Astronomy, State University of New York, Stony Brook,
\mbox{New York} 11794-3800; http://ms.cc.sunysb.edu/\~{}rmould}}  
\date{}    
   
\begin{document}             

\maketitle              

\begin{abstract}

The nRules are empirical regularities that were discovered in macroscopic situations where the outcome is known.  When they are
projected theoretically into the microscopic domain they predict a novel ontology including the frequent collapse of an atomic wave
function, thereby defining an nRule based foundation theory.  Future experiments can potentially discriminate between this and other
foundation theories of (non-relativistic) quantum mechanics.  Important features of the nRules are: (1) they introduce probability
through probability current rather than the Born rule,  (2) they are valid independent of size (micro or macroscopic), (3) they apply
to individual trials, not just ensembles of trials. (4) they allow all observers to be continuously included in the system
without ambiguity, (5) they account for the collapse of the wave function without introducing new or using old physical constants,
and (6) in dense environments they provide a high frequency of stochastic localizations of quantum mechanical objects.  Key
words: measurement, stochastic choice, state reduction.

\end{abstract}

\section*{Introduction}
The nRules are four auxiliary rules that guide or direct the application of Schr\"{o}d-inger's equation. They are verbal instructions
that say how the Schr\"{o}dinger equation should be applied to any quantum mechanical system.  One of them (nRule 4) modifies the
Hamiltonian in a way that will be shown in the next section. 

Every equation in physics is accompanied by verbal instructions of some sort.  Otherwise it would not be physics -Ð it would be
mathematics.  The nRules are different from most auxiliary rules in that they impose limitations on the Schr\"{o}dinger equation that
have no classical counterpart.  They are more assertive and have direct dynamical consequences.  

The Born rule is an auxiliary rule of standard quantum mechanics; however, it is not necessary as a governing tenet.  It can
be derived  from other rules, and the notion of probability can be introduced into quantum mechanics through
\emph{probability current}.  This has been done in two cases:  The \emph{nRules} \cite{RM1, RM2}  and the \emph{oRules}
\cite{RM3, RM4}.  All other quantum mechanical auxiliary rules that do not use probability current in this way will be referred to as
\emph{sRules}. This paper is primarily concerned with the nRules.

There are several consideration that go into the formulation of the nRules.  It is required that wave collapes or state reductions
are `objective' and `self-generated'.  That is, the rules describe how a wave function can collapse automatically inside a closed
system, independent of any outside observer or measuring device.  In addition, the rules are intended from the beginning to apply to
individual interactions (single trials) not just ensembles of trials, and to be independent of size.  It is initially decided to
use probability current to introduce probability.  And finally, the nRules are discovered by testing their validity in well-known
macroscopic situations where they can be shown to be empirically correct. They are then projected theoretically into the microscopic
domain.

As a consequence of these requirements it is found that the nRules allow the \emph{primary} observer to be continuously included in
the system.  This is similar to classical physics in that an observer who investigates an external system has the option of extending
the system to include himself.   The sRules do not always let that happen.  In the Copenhagen case the Born rule requires that the
primary observer remain outside.  He is allowed to peek at the system from time to time to determine the Born connection at that
moment, but he is not allowed to stay in the system.  Other sRules such as the many world thesis of Everett and the GRW/CSL
reduction theory of GianCarlo Ghirardi and his associates also allow the observer to be in the system in the classical sense
\cite{HE,GRW,GPR}.

Another consequence is that any and all secondary observers can be included in the system in a continuous and unambiguous way.  This
removes the paradoxical results that are associated with the Schr\"{o}dinger cat experiment, and with all other ambiguities that
result when a secondary observer is admitted into the system.  Therefore, \emph{all} conscious observers have a place in quantum
mechanical systems under the nRules.  

The above comments also apply to the oRules with the exception of the ``observer independence" of state reduction, inasmuch as the
oRules require that a conscious observer must be present for a measurement to occur -- following von Neumann's suggestion to that
effect.  An objection to the nRules is considered in the Conclusion together with a discussion of the experimental possibilities.

\section*{The nRules}
We define \emph{ready components} to be the basis components of state reduction.  These are the components that are chosen to survive
the collapse of the wave function. They are underlined throughout the paper.   Components that are not ready are called
\emph{realized components} and appear without an underline.  

\vspace{.4cm}

The first nRule describes how ready components are introduced into solutions of Schr\"{o}dingerÕs equation.  

\noindent
\textbf{nRule (1)}: \emph{If an irreversible interaction produces a complete component that is discontinuous with its predecessor
in some variable, then it is a ready component.  Otherwise a component is realized.}

\noindent
[\textbf{note:} A \emph{complete component} is one that includes all the (anti)symmetrized objects in the universe.  Each included
object is itself complete in that it is not a partial expansion in some representation.]

The second rule establishes the existence of a stochastic trigger.  The flow per unit time of square modulus is given by the
square modular current $J$, and the total square modulus of the system is given by $s$.

\noindent
\textbf{nRule (2)}: \emph{A systemic stochastic trigger strikes a ready component with a probability per unit time equal to the
positive probability current J/s flowing into it.  A realized component is not stochastically chosen.}

\noindent
[\textbf{note}: The division of $J$ by $s$ automatically normalizes the system at each moment of time.  Currents rather than
functions are normalized under these rules.]

The collapse of a wave is given by nRule (3)

\noindent
\textbf{nRule (3)}: \emph{When a ready component is stochastically chosen it will become a realized component, and all other
(non-chosen) components will go immediately to zero.}

\noindent
[\textbf{note}: We can choose to amend nRule (3) so that other components do \emph{not} go to zero.  It does no harm to let them
stand unchanged after a stochastic hit because there will be no further consequence.  Square modulus has no physical meaning in the
nRules, and current no longer flows into or out of these components because the Hamiltonian has passed them by -- as will be
explained in the next section.  The spent components would have the status of a ``phantom" as defined in Ref.\ 1.  The decision to
let these   not-chosen components go to zero or to let them stand is like the decision in standard quantum mechanics to renormalize
(or not) after a measurement]

The fourth nRule has a less obvious meaning.  

\noindent
\textbf{nRule (4)}: \emph{A ready component cannot transmit probability current to other components or advance its own
evolution.}

\noindent
[\textbf{note}: The fourth nRule is enforced by withholding a ready component's Hamiltonian as explained in the next section.]

\section*{The Quantum Algorithm}
To understand nRule (4) and the note under nRule (3), we go back to the Hamiltonian formalism and adopt a modification that I call
the  ``Quantum Algorithm".  Instructions for the application of the classical Hamiltonian are: \emph{Beginning
with the initial boundary conditions of a closed system, the Hamiltonian drives all of the system's particles and all of its
interactions \textbf{into the indefinite future}}.  This defines  a classical unitary evolution that goes on forever.

The quantum algorithm modifies these instructions to read:

\noindent
QUANTUM ALGORITHM: \emph{Beginning with the initial boundary conditions of a closed system, the Hamiltonian drives all of the
system's particles and all of its interactions  \textbf{up to but not beyond} the next ready component(s)}.  This introduces a
non-unitary process at each ready component that locates sites (or possible sites) of quantum measurement. It also locates other
non-unitary evolutions that are assumed to occur microscopically. 

The fourth nRule is implicit in this algorithm.  In fact, the quantum algorithm can be used in place of nRule (4).

 It works like this.

The horizontal line labeled $S_0$ in Fig.\ 1 represents the initial state of a system that evolves continuously in time.  It is a
single (complete) component driven by the Hamiltonian $H_0 + H_{01}$ until it encounters and includes the discontinuous
irreversible gap $G_{01}$.  The interaction Hamiltonian $H_{01}$ will drive probability current across this gap to the first
`ready' component in the figure (i.e., the first shaded area), but  current will not pass beyond that point.  As a result, the shaded
component in $G_{01}$ will accumulate square modulus.  This blocking of probability current (required by nRule 4)
 is accomplished by truncating the initial Hamiltonian so it does not contain the term $H_1$.

\begin{figure}[t]
\centering
\includegraphics[scale=0.8]{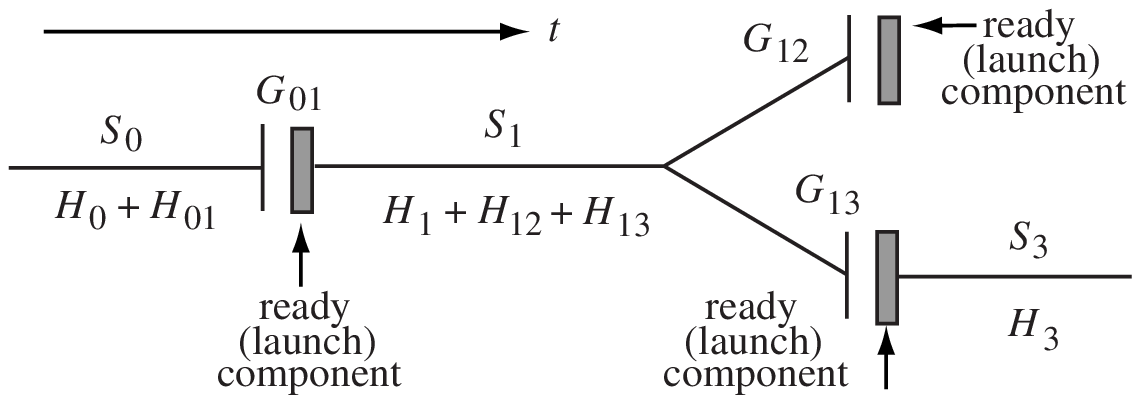}
\center{Figure 1: Collapse $S_0$ to $S_1$ then to $S_3$  }
\end{figure}

A stochastic hit on the ready component in $G_{01}$ will change it to a `realized' component according to nRules (3), thereby
launching a new solution $S_1$ of Schr\"{o}dinger's equation.  It will also make the initial system state irrelevant or  go to zero
(i.e., it \emph{reduces} $S_0$ to $S_1$).  The  ready component is also called a \emph{launch component}, for it contains all of the
\emph{initial conditions} of the launched solution $S_1$.  These initial conditions are continuously updated by current flow into the
launch component prior to the stochastic hit.  We therefore say that the collapse of the wave across $G_{01}$ is characterized by a
\emph{new solution that is brought about by a new Hamiltonian}, where the initial conditions are
contained in the launch component.  All of this can just as well be accomplished if, in place of nRule (4), the quantum algorithm is
directly applied; for in that case, the second Hamiltonian $H_1 + H_{12} + H_{13}$ will substitute for $H_0 + H_{01}$ the moment
initial conditions contained in $G_{01}$ are launched to give the new solution $S_1$.  The variables of $H_0$ and $H_1$ are indexed
to the variables of $S_0$ and $S_1$ respectively.

Once $S_1$ is launched, the new Hamiltonian will carry the system up to and across the next discontinuous and irreversible gap(s). 
Let this consist of the two parallel gaps $G_{12}$ and $G_{13}$ as shown in Fig.\ 1, so the launch components in  $G_{12}$ and
$G_{13}$ present the stochastic chooser with two possibilities.  The Hamiltonian driving $S_1$ is $H_1 + H_{12} + H_{13}$ as stated
above, where $H_{12}$ and $H_{13}$ are discontinuous interaction terms, so positive current will flow into both of the parallel ready
components.  However, only one will be chosen.  Suppose it is $S_3$ together with
Hamiltonian $H_3$ with everything else going to zero or just being ignored.  In Fig.\ 1 we conclude with this solution and its
Hamiltonian.  As in $G_{01}$, the flow of this system at both $G_{12}$ and$G_{13}$ is non-unitary.

\section*{A Particle Capture}
These nRules were initially discovered by examining many different quantum mechanical interactions that involve classical
instruments like detectors or counters.  They successfully describe all the cases considered; and on this basis, they are
claimed to describe \emph{any} observable quantum phenomena.  This section is the first example of this macroscopic kind, where an
elementary particle is captured by a detector.  There can be no question about the macroscopic events that occur as a result of that
interaction.  The nRules describe `what happens' in this case without any attempt to give a theoretical explanation, or to justify
those rules beyond the fact that they work. 

We apply Schr\"{o}dinger's equation to a particle interacting with a detector in order to demonstrate the first stage of Fig.\ 1. 
The interaction beginning at time $t_0$ is given by 
\begin{equation}
\Phi(t\ge t_0) = \psi(t)d_0(t) + \underline{d}_1(t)  \hspace{.5cm}\mbox{applying}\hspace{.1cm} H_0 + H_{01}
\end{equation}
where the second component is zero at $t_0$ and increases in time.  The free particle $\psi(t)$ here interacts with the ground state
detector $d_0(t)$ and is driven by the Hamiltonian $H_0$.  The interaction Hamiltonian $H_{01}$ produces a probability current flow
from the first component to the second component in Eq.\ 1, where the latter is the detector in its capture state.  The gap between
these two components is discontinuous because the particle is completely outside the detector in the first component, and it is
completely inside the detector in the second component; and the two are not bridged by intermediate components that are continuous in
particle variables.  This interaction is also irreversible.  Therefore, the gap given by the + sign satisfies nRule (1), making
$\underline{d}_1(t)$ a `ready' component as indicated by the underline.  Each component in Eq.\ 1 is multiplied by the associated
total environment (not shown), assuring detector decoherence in this case and satisfying the requirement that each component is
complete.  

Current flowing into $\underline{d}_1(t)$ will increase its square modulus and update its content through the interaction Hamiltonian
$H_{01}$, but any further evolution is blocked by nRule (4).  This makes $d_1(t)$ the launch component that establishes and updates
the initial conditions of the next solution of the Schr\"{o}dinger equation.  Since positive probability current flows into this
ready component, it is subject to a stochastic hit as specified by nRule (2).  If that happens at a time $t_{sc}$, then nRule (3) will
require a state reduction giving 
\begin{equation}
\Phi(t\ge t_{sc} > t_0) = d_1(t)  \hspace{.5cm}\mbox{now applying}\hspace{.1cm} H_1 
\end{equation}
At $t_{sc}$ the Hamiltonian $H_1$ is applied to the realized detector component $d_1(t)$, allowing it to evolve on its own. 

 The time dependence of $d_1(t)$ in Eq.\ 2 refers to the evolution of the detector after capture.  The time dependence of
$\underline{d}_1(t)$ in Eq.\ 1 refers to changes in the launch component due to current flow from the first component.  The latter
changes include the increase in the square modulus of the launch, plus information that continuously updates the boundary conditions
that are contained in the launch.  The distinction between these two changes will be better clarified in next section.  The necessity
of nRule (4) will become more apparent in subsequent sections where it is shown that a macroscopic body (such as a counter) cannot be
otherwise described by Schr\"{o}dinger's equation in this non-Born rule protocol.

\section*{Free Neutron Decay}
When the nRules are applied to microscopic systems they become `speculations' rather than empirically knowable regularities.  This
section is the first example of that kind.  The nRules are here projected into a realm in which the resulting ontology is very
different from the description given by standard quantum mechanics.  Again, there is no attempt to justify the nRules beyond the fact
that they work well on the macroscopic level.  Our speculation amounts to requiring that there is no fundamental distinction between
the macro and the microscopic, so that the same rules -- in particular the nRules -- apply in both domains.  

A \emph{free neutron decay} is  given by
\begin{displaymath}
\Phi(t \ge t_0) = n(t) + \underline{e}p\overline{\nu}(t)
\end{displaymath}
where the second component is zero at $t_0$ and increases in time.  It is shown as a package of three particles that are the boundary
conditions of the neutron's decay.  It's a ready component, although it not necessary to underline the entire component -- one state
will do.   To satisfy the requirement of completion, each component is multiplied by the total
environment (not shown) even though the neutron and its decay products are an isolated system.

This case provides a good example of how the launch component $\underline{e}p\overline{\nu}(t)$ is a function of time
beyond its increase in square modulus. Assume that the neutron moves across the laboratory in a wave packet of finite width.  At each moment the launch component will ride
\emph{with the packet}, having its same shape and group velocity.  This component contains the boundary conditions of the next
solution of the Schr\"{o}dinger equation -- the solution that appears when $\underline{e}p\overline{\nu}(t)$ is stochastically chosen
at $t_{sc}$.  The launch component is time dependent because it increases in square modulus \emph{and} because it follows the motion
of the neutron.  This is how it updates the boundary conditions of the decay.  However, nRule (4) insures that the launch will not
evolve dynamically beyond itself before becoming a realized component at the time of stochastic choice.  Only then will the neutron
disappear by decaying into separate particles $e(t)p(t)\overline{\nu}(t)$that spread out on their own, still correlated in conserved
quantities.  

As a consequence of nRules (3) and (4), the decay occurs at only one stochastically chosen point along the path; whereas, standard
quantum mechanics results in a superposition of decays that are spread out all along the path.  The nRules apply to individual
locations along the path, and the sRules apply to ensembles of those locations.  This shows that the nRules are more definite than
the sRules, although they remain less deterministic than classical physics.

Specific values of the electron's momentum are not stochastically chosen by this reduction.  All the possible values of momentum are
included in $e(t)p(t)\overline{\nu}(t)$ after the transients associated with $\Delta E\Delta t$ have died out, where $\Delta t$
begins only after the new Hamiltonian is applied.  For the electron's momentum to be determined in a specific direction away from the
decay site, a detector in that direction must be activated.  That will require another nRule equation involving a stochastic
hit on the detector. 

Experimentally it should be possible in a single trial to find the momentum of the electron, the proton, and the antineutrino with
sufficient accuracy to locate the decay somewhere along the original path of the neutron with an accuracy consistent with
Heisenberg.  However, this information cannot be used to disqualify the nRule or the standard theoretical description of what
happened.  Experiment cannot confirm or deny the existence of intermediate boundaries (or wave reductions) that fix the locus the
decay before the separate particles have been detected.  In this and many other cases, our theoretical projection of the nRule gives
us a microscopic ontology that is unique, but it is experimentally indistinguishable from that of standard quantum mechanics.

\section*{A Series of Discontinuities}
The section on ``Particle Capture" allowed the nRules to be given their first macroscopic formulation.  The next macroscopic step is
to consider a series of discontinuities as represented by a particle counter $A$ that is activated by a nearby radioactive source.

Consider the series of components $A_0$, $A_1$, $A_2$, $A_3$, ...
that are serially connected to each other by discontinuous and
irreversible gaps.  In standard quantum mechanics, a series like this is given by
\begin{equation}
\Phi(t \ge t_0) = A_0(t) + A_1(t) + A_2(t) + A_3(t) + ...
\end{equation}
where only $A_0$ is non-zero at time $t_0$.  The other components gain amplitude by virtue of probability current flowing from $A_0$
to $A_1$, then to $A_2$, and then to $A_3$, etc.  They are a succession of counter readings whose subscripts record the number of
particles that have been captured from a nearby radioactive source.  Let $A_0$ mean that no particles have been captured, $A_1$ means
that one particle has been captured, etc.  We do not include the intermediate particle field in \mbox{Eq.\ 3}.  It  simplifies the
example to imagine that the apparatus states interact directly with each other.

Applying the \emph{first three} nRules to this case, the envelope of the amplitudes of the components in Eq.\ 3 will form a pulse that
moves from left to right.  All the current receiving components will be ready components according to \mbox{nRule (1)}, so \mbox{Eq.\
3} will take the form
\begin{equation}
\Phi(t \ge t_0) = A_0(t) + \underline{A}_1(t) + \underline{A}_2(t) + \underline{A}_3(t) + ...
\end{equation}
But there is a problem.  In the absence of nRule (4), the component $\underline{A}_2(t)$ will acquire some degree of amplitude the
moment $\underline{A}_1(t)$ acquires amplitude.  Probability current might then flow simultaneously into $\underline{A}_1(t)$ and
$\underline{A}_2(t)$, in which case there might be a stochastic hit on $\underline{A}_2(t)$ before there is a hit on
$\underline{A}_1(t)$.  That is a very unphysical result for a macroscopic counter.  The fourth nRule is added to insure that
\emph{this does not occur}.

When nRule (4) is applied, Eq.\ 4 becomes
\begin{equation}
\Phi(t \ge t_0) = A_0(t) + \underline{A}_1(t) 
\end{equation}
and this guarantees that the state $\underline{A}_2(t)$ will not be stochastically chosen before $\underline{A}_1(t)$.  The effect
of nRule (4) is therefore to guarantee that $\underline{A}_1(t)$ is not passed over, and this alone is an indispensable requirement
in this non-Born protocol. It is a macroscopic necessity that alone justifies our adopting the truncated Hamiltonian of the quantum
algorithm.

Probability current flowing from $A_0$ to $\underline{A}_1$ in Eq.\ 5 will result in a stochastic hit on $\underline{A}_1$ at some
time $t_{sc1}$.  When that happens we get the first particle capture
\begin{equation}
\Phi(t \ge t_{sc1} > t_0) = A_1(t) + \underline{A}_2(t) 
\end{equation}
where $\underline{A}_2(t)$ is zero at $t_{sc1}$ and increases in time.  Following this, another stochastic hit at $t_{sc2}$ gives the
second particle capture

\pagebreak

\begin{equation}
\Phi(t \ge t_{sc2} > t_{sc1} >t_0) = A_2(t) + \underline{A}_3(t) 
\end{equation}
and so fourth.  In Eqs.\ 5, 6, and 7, the correct sequential order of counter states is guaranteed by nRule (4).

It is characteristic of the sRules (i.e., any Born-based theory) that there is only one solution (Eq.\ 3) to the Schr\"{o}dinger
equation for the given initial conditions, whereas the nRules provide a separate solution for each discontinuous gap (Eqs.\ 5, 6, 7,
etc.).  The launch component in each case will contain all of the updated boundary conditions of the next solution that become
effective the moment it is stochastically chosen.  So $\underline{A}_1$ in Eq.\ 5 is the launch component into the new solution in
Eq.\ 6 and contains all of the boundary conditions that apply at time $t_{sc1}$.  The realized component $A_1(t_{sc1})$ in Eq.\ 6 is
the initial boundary of that solution.  

There is no contradiction between the predictions of the nRules and the standard sRules.  The nRules are concerned with the
probability in an individual trial that the next stochastic hit will occur in the next interval $dt$ of time.  Opposed to this, the
sRules are concerned with the distribution of an \emph{ensemble} of states at some finite time $T$ after the apparatus is turned on. 
These different rules-sets ask different questions having different answers.  However, either one of these protocols can be
successfully mapped onto the same counter, so there can be no observational contradiction. 

It is no strain to see that Eq.\ 5 applies to \emph{microscopic states} as well, for serial order is just as important in these
cases.   Atomic states that decay from an initial excited state $A_0$ will go to the next lower energy state $A_1$ and then lower
to $A_2$ without skipping a step -- unless that possibility is allowed by the Hamiltonian.  If it is not allowed, then
$A_1$ will not be skipped over.  As in the macroscopic case, nRule (4) is an essential moderator of any serial sequence
at the atomic level.  Otherwise, the second order component $A_2$ might be stochastically chosen before
$A_1$ is chosen, and that would be unphysical.  Although the nRules are empirically discovered by investigating
macroscopic systems, they can be extended to this microscopic system, thereby supporting the claim that the rules are independent of
size.

\section*{Parallel Discontinuities}
Macroscopic parallel branching is also used to check the correctness and generality of the nRules.  The equation of state involving a
laboratory apparatus $A$ is given by
\begin{equation}
\Phi (t \ge t_0) = A_0(t) + \underline{A}_r(t) + \underline{A}_l(t)
\end{equation}
where $\underline{A}_r(t)$ and $\underline{A}_l(t)$ are the eigencomponents of stochastic choice that are initially equal to zero
and increase in time.  Each one receives probability current from the first component that makes it a candidate of a state reducing
stochastic hit -- like $G_{12}$ and $G_{13}$ in Fig.\ 1.  Each is a launch component, where $\underline{A}_r(t)$ contains the
boundary conditions of a launch toward the right in Fig.\ 2, and $\underline{A}_l(t)$ contains the boundary conditions of a launch
toward the left.  The dashed lines in Fig.\ 2 are the initially forbidden transitions.

\begin{figure}[b]
\centering
\includegraphics[scale=0.8]{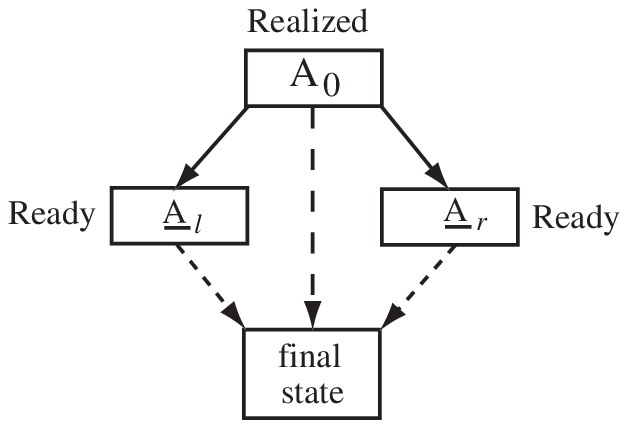}
\center{Figure 2: Possible parallel decay routes}
\end{figure}

Think of these components as representing two macroscopic particle counters, where $A_0$ means that neither one has yet made a
capture, $A_r(t)$ means that the one on the right is the first to make a capture, $A_l(t)$ means that the
one on the left is the first to make a capture, and the final state $A_f(t)$ represents the system when each counter has made a
single capture.  Let each counter turn off after a single capture.  Again, we simplify by not including the particle fields.  If the
launch component $\underline{A}_r(t)$ is stochastically chosen in Eq.\ 8 at time $t_{scr}$, the resulting state reduction will yield
\begin{displaymath}
\Phi (t\ge t_{scr} > t_0) = A_r(t) + \underline{A}_f(t)
\end{displaymath}
where $\underline{A}_f(t)$ is the launch component into final state of the system.  When it is stochastically chosen at time
$t_{scf}$ the system will be in its final state \mbox{$\Phi (t\ge t_f > t_{scr} > t_0) = A_f(t)$}.  This sequence will go in a
counterclockwise direction if the launch component $\underline{A}_l(t)$ in Eq.\ 8 is stochastically chosen.  As in the series case,
all of these launch components are time dependent because their square moduli increase in time \emph{and} because at each moment
they take on the updated boundary conditions that apply to the new solution in case of a stochastic hit at that moment.

The fourth nRule therefore has the effect of forcing these macroscopic counters into either a clockwise or a counterclockwise path in
the classical sense.  Without nRule (4), a second order transition might skip over the intermediate components to score a direct
stochastic hit on $A_f$ without one of the intermediate component being definitely involved.  This is unphysical behavior for a
macroscopic system.  So nRule (4) transforms the initial superposition of \mbox{Eq.\ 8} into two classical alternatives because it
does not allow intermediate components such as $\underline{A}_l$ or $\underline{A}_r$ to be skipped over.  Here again we see the
indispensability of nRule (4) if macroscopic objects are to be quantum mechanically described with a non-Born protocol.

The same will be true of \emph{microscopic} parallel systems.  An ``irreversible discontinuity" imposes an abrupt and lasting change
of a distinctive kind in some part of the universe -- even in a microscopic case.  For instance, let Fig.\ 2 represent two alternative
routes from a high-energy atomic state $A_0$ to the ground state $A_f$.  The two photons that are released along each path will leave
an irreversible record that will be different for each path (assuming non-degeneracy); so if the two photons associated with the
clockwise path are found in the wider universe, then the clockwise path must have been stochastically chosen.  It is not possible for
\emph{all four} photons to be found in a single trial.  It will be either the two photons from the left or the two from the right.  The
released photons are the abrupt and lasting change referred to above, and the distinctive characteristics of the photons along each
path removes the possibility of interference between the paths. Statistically, the two paths are a mixture; so in any individual
trial, only one eigenstate $\underline{A}_r$ or $\underline{A}_l$ would be traversed.  

More generally of any microscopic or macroscopic series/parallel combination of paths, any single path segment that follows and
precedes an irreversible discontinuous gap will be phase independent of all the other such path segments in the combination, and will
be correctly described by the nRules.

\section*{Add an Observer}
When an observer is added to the system it is necessarily macroscopic.  The results of this section are empirically undeniable, and
may therefore be considered a further check on the correctness of the nRules.

First, imagine that an observer is present to witness the  detector capture in Eq.\ 1 as it would appear following any one of the
sRules of standard quantum mechanics.   The resulting equation would be 
\begin{equation}
\Phi (t \ge t_0) = \psi d_0B_0 + d_{1w}B_0 \rightarrow d_{1d}B_1
\end{equation}
where only the first component is non-zero at $t_0$ and decreases in time.  To simplify, the time dependence of the states in these
components is not explicitly shown.  The  state $d_{1w}$ represents the detector when the captured particle has advanced to
`just inside' the \emph{window}.  This component will evolve continuously from that point on, carrying the influence
of the capture from the window end of the detector to the \emph{display} end, which is given by the display  state
$d_{1d}$.  The arrow in Eq.\ 9 represents this continuous evolution.  So $d_{1w}B_0 \rightarrow d_{1d}B_1$ is a single component of
Schr\"{o}dinger's equation that evolves in time, representing the continuous/classical change that occurs inside the detector after a
capture.  

The state $B_0$ in Eq.\ 9 is the brain state of the observer who witnesses the detector in its ground state.  The brain will continue
in that state after capture until the signal has moved through the detector to the display.  At the display end it becomes $B_1$,
which is the brain state of the observer who witnesses the capture.  The change from $B_0$ to $B_1$ is therefore continuous and
largely classical.  The quantum mechanical discontinuity is confined to the particle's jump when it goes from being outside of the
detector to being just inside the window.  Of course the detector also jumps discontinuously from $d_0$ to $d_{1w}$ at the same time.

The trouble with Eq.\ 9 is that it is a potential cat-like disaster.  If the interaction time between $d_0$ and the particle field is
long enough (i.e., if the incoming particle is spread out sufficiently in space), the initial signal will travel through to the
display before the first component has gone to zero.  This means that two very different brain states $B_0$ and $B_1$ will appear
simultaneously in that equation, and that will result in an ambiguity reminiscent of Schr\"{o}dinger's cat experiment. 

When the nRules are applied to this case, Eq.\ 9 is modified to
\begin{equation}
\Phi (t \ge t_0) = \psi d_0B_0 + \underline{d}_{1w}B_0 
\end{equation}
where $\underline{d}_{1w}B_0$ is the launch component of the gap that contains the boundary conditions of the next solution.  This
component is equal to zero at $t_0$ and increases in time.  Again, it is sufficient to underline only one state in the launch
component to indicate that the entire component is `ready'.  Following a stochastic hit on this component at time $t_{sc}$, we get
the next solution
\begin{equation}
\Phi (t \ge t_{sc} > t_0) = d_{1w}B_0 \rightarrow d_{1d}B_1 
\end{equation}
where the resulting realized component is $d_{1w}B_0$ at $t_{sc}$ and evolves continuously in time to become $d_{1d}B_1$.

There is no ambiguity here because there is only one brain state in Eq.\ 10, and there is only one brain state `at a time' in Eq.\
11.  The Copenhagen rules give us one equation (Eq.\ 9) in which there is a potential cat-like ambiguity, whereas the nRules give us
two equations (Eqs.\ 10 and 11) in which there is no ambiguity.  Equation 10 applies \emph{before} a stochastic hit, and Eq.\ 11
applies \emph{after} a stochastic hit.  Here again, the standard sRules compress two solutions having two separate boundary conditions
into a single equation with only the initial conditions $d_0B_0$, so these solutions are \emph{not properly grounded} in all of the
boundary conditions that apply.   On the other hand, the nRules give two solutions, each based on different boundary conditions given
by
$d_0B_0$ and
$d_{1w}B_0$.

In standard sRule theories, the Born rule provides the probability connection between theory and observation through the square
modulus.  The nRules establish this connection differently.  An observer's brain is placed in the system in contact with the
apparatus (e.g., a detector or counter), and his experience is predicted by the effect of that interaction on the brain.  The
probability of this happening during a given time interval $dt$ is determined by the probability current.  As in classical physics,
the primary observer can imagine that his own brain is part of the system.  This means that the primary and secondary observers have
the same ontological status under the nRules, and that both have the same `reality', as any other object in the universe.  The
primary observer is not banished to an Olympian mountaintop where he studies the distant universe as something apart from himself.

\section*{Multiple Parallel Sequences}
Construct a network of macroscopic counters and sources that allow an initial state $AB_0$ to decay to either eigencomponents
$AB_1$ or $AB_2$ or $AB_3$, where $A$ is an apparatus that is witnessed by a brain state $B$.  The subscript on $B$ denotes the state
of the apparatus that is seen by the brain.  Ignore the window/display distinction, and again assume that the components are time
dependent.  After the first stochastic choice that carries the system from the first to the second row of Fig.\ 3, there is a
secondary choice that carries the system to the third row. 

\begin{figure}[t]
\centering
\includegraphics[scale=0.8]{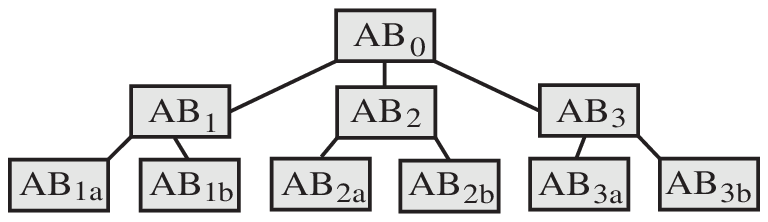}
\center{Figure 3: Six possible sequences }
\end{figure}

There are six possible sequences in this diagram.  In the many-world universe of Everett these branches all run together in a single
superposition.  The observer who inhabits one of these branches cannot be aware of his own alter-ego in another branch, for that
would disqualify the idea.  Everett showed that once begun, one of these sequences will proceed without any further involvement with
any other sequence.  That means that the observer on one branch of this system will not be aware of his alter-ego on another branch. 
The branches (or sequences) therefore proceed independent of one another, even though they are regarded (by Everett) as a single
superposition.  As before, the particle field is not included in the analysis.

The nRules tell a different story.  They do not run all these solutions together.  The nRules say that each stochastic choice in
Fig.\ 3 is the occasion of a collapse of the wave and the launch of a new solution of Schr\"{o}dinger's equation.  Each of the six
possible sequences consists of two collapses that follow the initial state $AB_0$.  There will therefore be three separate equations
that carry the initial state into a final state.  For the sequence $AB_0$, $AB_1$, $AB_{1b}$, those equations are
\begin{eqnarray}
\Phi(t \ge t_0) &=& AB_0 + \underline{A}B_1 + \underline{A}B_2 + \underline{A}B_3 \nonumber  \\
\Phi(t \ge t_{sc1} >t_0) &=& AB_1 + \underline{A}B_{1a} + \underline{A}B_{1b} \nonumber  \\
\Phi(t \ge t_{sc2} > t_{sc1} >t_0) &=& AB_{1b}  \nonumber
\end{eqnarray}
where in each case the launch components are initially zero.  Evidently the system is not a superposition of all the possible
sequences; but rather, each sequence is a series of individual decays that proceed independent of other sequences. 

Although this construction is illustrated with observable macroscopic instruments, it would work as well with a microscopic array of
atomic states where there can be no observers.  This again is because every irreversible \& discontinuous gap and accompanying
stochastic hit leaves a mark on the wider universe that indelibly records the choice.  In the microscopic case this mark will take
the form of an emitted photon or other irreversible happening recorded in the `memory' of the universe, much as the memory of each
alter-ego is irreversibly affected in Everett's theory.

\section*{Atomic Absorption and Emission}
Applying this scheme to the case of atomic absorption and emission, the atom in its ground state interacts with a laser field
$\gamma_n(t)$ containing \emph{n} photons of the excitation frequency.  These are photons of frequency 0-1, where 0 refers to the
ground state $a_0(t)$, and 1 refers to the excited state $a_1(t)$.  The nRules then give
\begin{equation}
\Phi (t \ge t_0) = \gamma_na_0 \Leftrightarrow  \gamma_{n-1}a_1 + \gamma_{n-1}\underline{a}_0 \otimes \gamma
\end{equation}
where only the first component is zero at time $t_0$.  Again, only one state in the ready component $\gamma_{n-1}\underline{a}_0
\otimes \gamma$ needs to be underlined, and it is understood that every state is a function of time.  The double arrow
$(\Leftrightarrow)$ represents a reversible Rabi oscillation associated with the laser that begins at $t_0$.  When the atom is in the
excited state $a_1$ a spontaneous emission to ground becomes a possibility, represented here by the ready component.  When that
component is stochastically chosen the atom goes to ground, emitting a photon $\gamma$ that came to it from the laser beam.

If the atom begins in the excited state and is exposed to a laser beam, we get
\begin{equation}
\Phi (t \ge t_0) = \gamma_na_1 \Leftrightarrow  \gamma_{n+1}a_0 + \gamma_{n}\underline{a}_0 \otimes \gamma
\end{equation}
where again, only the first component is non-zero at $t_0$.  Again, a stimulated emission oscillation begins immediately, where a
spontaneous emission from the excited state is represented by the ready component.  Except for the fact that Eq.\ 13 has one more
 photon than Eq.\ 12, the two equations are identical.  It cannot matter if the oscillation begins in $a_0$ or in $a_1$.

\section*{A Laser}
Given a four level atom with a ground state $a_0$ and three excited states $a_1$, $a_2$, $a_3$ of increasing energy.  It is immersed
in a laser field of $n$ photons $\gamma$ with an energy that connects levels $a_1$ and $a_2$.  The atom is initially pumped into the
short-lived state $a_3$ and is quickly dropped into $a_2$ by an irreversible energy loss involving some dissipative process that may
be molecular collisions, or possibly the spontaneous emission of a 3-2 photon.  
\begin{equation}
\Phi (t \ge t_0) = \gamma_na_3 + \gamma_n\underline{a}_2\otimes e_x
\end{equation}
where the second component is zero at $t_0$ and increases in time.  The symbol $e_x$ represents other parts of the system not
appearing in the first component (like adjacent molecules or radiation fields) that absorb the energy difference.  With a stochastic
hit on the ready component in this equation at time $t_{sc1}$, the system becomes
\begin{eqnarray}
\Phi(t \ge t_{sc1} > t_0) &=& \gamma_na_2\otimes e_x \Leftrightarrow  \gamma_{n+1}a_1\otimes e_x + \gamma_n \underline{a}_1 \otimes
e_x \otimes \gamma \hspace {.2cm}\mbox{(metastable)}\nonumber  \\
 &&  \hspace {4.1cm} + \hspace {0.1cm} \gamma_{n+1} \underline{a}_0 \otimes e_x \otimes e_{xx}\nonumber  
\end{eqnarray}
where only the first component is non-zero at $t_{sc1}$.  The metastable decay in the first row is a long-lived spontaneous photon
emission coming off the first component.  The symbol $e_{xx}$  in the short-lived decay component (second row) represents that part
of the environment that takes up the energy difference between $a_1$ (in the second component) and $a_0$.  The short-lived decay
product is more likely to be stochastically chosen than the metastable one, so after a second hit at time $t_{sc2}$ we have
preferentially
\begin{displaymath}
\Phi(t \ge t_{sc2} > t_{sc1} > t_0) = \gamma_{n+1}a_0 \otimes e_x \otimes e_{xx}
\end{displaymath}
Comparing the original state $\gamma_na_3$ with the final state $\gamma_{n+1} a_0 \otimes e_x \otimes e_{xx}$ , it is
clear that the energy difference between $a_3$ and $a_0$ is the energy of the new photon in the laser beam plus the two dissipative
processes $e_x$ and $e_{xx}$. The cycle is repeated many times resulting in pumping many new photons into the laser beam. Evidently
each photon pumped into the beam requires two stochastic hits -- i.e., two wave collapses associated with two non-unitary processes. 
I do not call these 	``measurements" because I think it is best to reserve that word for non-unitary processes that involve
macroscopic instruments.

\section*{Localization}
Localization is essential if macroscopic objects are to have the location properties that correspond to our common experience with
them. This property is not contained in the Schr\"{o}dinger equation by itself, for objects subject only to that equation will expand
forever due to their uncertainty in momentum.  Therefore, localization must be provided for by auxiliary rules of some
kind.  The Copenhagen sRules use a macroscopic instrument to locate a quantum mechanical particle, but this will not apply when
macroscopic encounters do not play a fundamental role.  There is also the `bootstrap' question of how a macroscopic instrument can
itself be located, considering that it (and all other such instruments) began their existence as a collection of free hydrogen
and helium atoms following recombination about 14 billion years ago.  The Schr\"{o}dinger equation cannot itself localize (i.e.,
collapse) a collection of this kind.  On the other hand, the nRules make no essential use macroscopic objects.  These rules are shown
below to provide a purely microscopic localization of matter in certain matter-dense environments. 

Let a photon raise an atom to an excited state, after which the atom drops down again by spontaneously emitting a photon.  The
incoming photon is assumed to be spread out widely over space.  The atom is also spread out over space by an amount that exceeds its
\emph{minimum volume}.  This is defined to be the smallest volume that the atom can occupy consistent with its initially given
uncertainty of momentum.  The atom in Fig.\ 4 (shaded area) is assumed to be spread far beyond this volume prior to its interaction
with the photon. As the incoming photon passes over the enlarged atom, we assume that the scattered radiation will appear as a
superposition of many photons that originate from different parts of the atom's extended volume as shown in Fig.\ 4 -- these are the
small wavelets in the figure.  The correlations between the nucleus of the atom and the orbiting electrons must be preserved, even
though the atomic superposition is spread out over a much larger volume.  That is, the smaller dimensions of the minimal volume atom
must be unchanged during its expansion, so the potential energy of the orbiting electrons is unchanged.  The atom could not otherwise
act as the center of a `characteristic' photon emission.  This means that the incident photon will engage the compact atom throughout
every part of the enlarged volume.  

\begin{figure}[t]
\centering
\includegraphics[scale=0.8]{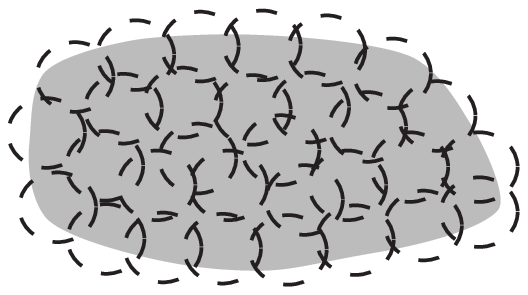}
\center{Figure 4: Scattered wavelets }
\end{figure}

However, if the atom is in a rich and random environment its volume in \mbox{Fig.\ 4} will not remain coherent.  Imagine that
external influences break it up into $n$ $bubbles$ of lesser volume (but no smaller than the minimum volume) that are decoherent
relative to one another.  We will then have
\begin{displaymath}
\Phi(r, t_{de} \ge t > t_0) = a(r, t) \rightarrow \Sigma_na_n(r, t)
\end{displaymath}
by the continuous decoherent process described in Ref.\ 1 where $t_{de}$ is a time when decoherence is complete.  In this equation
the initially coherent (shaded) atomic state $a(r, t)$ becomes $n$ separate decoherent bubbles, where each is represented by
$a_n(r, t)$.  This summation is not an expansion representation of the atom.  Rather, it is a mixture of decoherent
components, each of which is `complete' because the atom is entirely contained in each bubble.  When the incoming photon $\gamma$ is
included the above equation is
\begin{displaymath}
\Phi(r, t_{de} \ge t > t_0) = \gamma(t) a(r, t) \rightarrow \gamma(t)\Sigma_na_n(r, t)
\end{displaymath}

Let the photon interact with each of the bubbles after time $t_{de}$.  Each will then give rise to a Rabi oscillation between the
ground state $a_n(r, t)$ and the excited state given by \r{a}$_n(r, t)$.
\begin{displaymath}
\Phi(r, t \ge t_{de} > t_0) = \Sigma_n[\gamma(t) a_n(r, t) \Leftrightarrow  \mbox{\r{a}}_n(r, t)
+ \underline{a}_n(r, t)\otimes
\gamma_n(t)]
\end{displaymath}
where $\gamma_n$ is the photonic wavelet (in Fig.\ 4) associated with the $n^{th}$ bubble.  There is a chance that the $k^{th}$ ready
state in the square bracket will be stochastically chosen at a time $t_{sc}$.  In that case
\begin{displaymath}
\Phi(r, t \ge t_{sc} > t_{de} > t_0) = a_k(r, t)\otimes \gamma_k(r, t)
\end{displaymath}
When this happens the \emph{ground state atom is reduced} to the size of the $k^{th}$ bubble.  This illustrates one of the ways that
the world around us becomes localized.  Solar photons and greenhouse photons in the Earth's atmosphere will produce widespread
scattering events of this kind that localize the atmosphere, and hence the correlated surface of the earth.  The same will be true of
the scattering of the sun's light as it enters the waters of the Earth's lakes and oceans.  In addition, there are many examples of
irreversible reactions resulting from discontinuous quantum jumps in the rich biosphere of the earth -- all contributing to our
experience of being well defined in space.  A localization of this kind is discussed in another paper \cite{RM5}.

The above reduction is not possible under standard sRules because it does not involve a macroscopic object.  An exception would seem
to be the GRW/CSL theory because in this case a reduction to one of the bubbles is possible.  However, it is very unlikely.

\section*{Other Applications}
The nRules have been satisfactorily applied to a number of different macroscopic cases.  For instance, the Born rule can be `derived'
when an observer looks at the terminal results in a typical physics experiment.  In this case an interaction in an initially
normalized system is allowed to go to completion.  If there is more than one resulting launch eigencomponent, it follows from the
nRules that one of them will be chosen with a probability equal to the square modulus of that component -- which is the time integrated
probability current into that component.

The nRules also give good results when we investigate how an initially independent observer engages a particle detector
\emph{during} its interaction with a particle (Ref.\ 1).  Also in Ref.\ 1, we look at the case of two observers, where one joins the
other during an observed interaction between a particle and a detector.  

In a separate paper the Schr\"{o}dinger cat experiment is examined in all of its variations \cite{RM6}.  In one version the cat is
initially conscious and is made unconscious by a mechanical device that is initiated by a radioactive emission.  In another version
the cat is initially unconscious and is made conscious by an alarm clock that is set off by a radioactive emission.  In still another
version, the cat is awakened by a natural internal alarm (such as hunger) that is in competition with an external mechanical alarm. 
In all these cases, the nRules are shown to accurately and unambiguously predict the expected experience of the cat at any moment of
time.  And finally, an external observer is assumed to open the box containing the cat at any time during any one of these
experiments; and when that happens, his experience of the cat's condition is correctly predicted by the nRules.

In all the above macroscopic cases, plus the five examined in this paper, the nRules are found to be entirely correct.  Furthermore,
no macroscopic case has been found in which they are not correct.  It is easy to believe that rules that appear to be so generally
true at the macroscopic level are also true at the microscopic level; and when applied microscopically, the nRule description of
events is qualitatively different from that given by any of the sRules.  Several microscopic applications have been described in this
paper: neutron decay, series and parallel microscopic discontinuities, microscopic multiple sequences, atomic absorption and
emission, lasing, and localization.  Others are investigated in Ref.\ 1 where we look at spin states, decoherence, and Rabi
oscillations, neither of which brings about a state reduction under the nRules.  

There is an unexpected bonus contained in the nRules.  In another paper \cite{RM7}, nRule (4) insures the forward flow of
probability current.  In thermodynamics the forward direction is only very probable; but the nRules \emph{require} that time's arrow
flows in the right way.  Since nRule (1) contains the word ``predecessor", it must be reworded to say that the side of the gap with 
the higher entropy identifies the ready component.  With that change, it is clear that nRule (4) will prevent current from arriving 
at the gap from the ``wrong" direction, because that would require current flow within the ready component.

\section*{Conclusion}
One objection to these nRules is their verbal form. To be taken seriously it is said that a theory of physics must be formulated
mathematically.  For instance, \mbox{nRule (2)} requires the existence of a systemic stochastic trigger.  The GRW/CSL theory has a
similar requirement calling for the existence of a white noise or randomly fluctuating field $\omega (x, t)$.  There are some
functional differences between the two, like the selectivity of the trigger that only affects ready components, as opposed to the
noise that affects everything.  But the main objection to the trigger is that it is not represented by a mathematical function that
can be included in the equation of motion as can $\omega (x, t)$.  This, I believe, is an aesthetic objection that might or might not
be correct.  We've come to think of physical theory as being only mathematically expressible, but that might be too limited a view. 
Verbal rules, auxiliary to a dynamical principle, are no less precise, or true, because of their form.

A preference for the nRules is also an aesthetic choice at this point, for there is no experiment that can now distinguish it from
other foundation theories.  That may happen, but for the time being we are left with a three-step methodology that takes us from the
empirical to the theoretical and back to experimental possibilities.  

\noindent
\textbf{1}. The nRules are auxiliary rules of the non-relativistic Schr\"{o}dinger equation. They are empirically correct in all the
macroscopic situations that have been investigated.  I believe these investigations are sufficiently complete to claim that the
nRules are generally valid empirical laws or formulas - like Balmer's spectral series, or Planck's blackbody radiation law.

\noindent
\textbf{2}. It is easy to imagine that wide-spread regularities that appear in one domain will also appear in another domain.  If that
is true in this case, then the nRules will apply in microscopic systems as well as in macroscopic systems.  We assume here that the
fundamentals are the same in both domains, so both large and small things follow the same rules -- the nRules.  

\noindent
 \textbf{3}. The question is: Does experimental evidence exist that favors this theory compared with other current theories of quantum
mechanics and quantum measurement?  At present the answer is ``no".  However, there are prospects.  The GRW/CSL theory predicts the
existence of a physical constant $\lambda$ that governs the rate at which a particle is affected by the stochastic noise (Ref.\ 7). 
This is supposedly a very small number whose existence has not yet been experimentally confirmed.  If it is found at some future
time, then the microscopic generality of the nRules will no longer be defensible.  On the other hand, if experiments do not confirm
the existence of the GRW/CSL physical constant then all rival theories will remain viable candidates, including the microscopic
nRules.  I know of no experiment that can decisively verify the microscopic nRules, but I am hopeful that one will be found.

The more immediate virtue of the nRules is their heuristic value when thinking about microscopic processes.  The rules are no help
when calculating probabilities, but they do provide a skeletal outline that connects a microscopic network of stochastic choices,
making the options more transparent and the ontology better defined.  So in addition to the other desirable properties enumerated in
this paper (as in the abstract), the nRules help one understand why the Schr\"{o}dinger equation does what it does microscopically. 
Of course they are not just a heuristic device.  They claim to be an account of what really happens to atomic systems.


\begin{thebibliography}{99}


\bibitem{RM1}R. A. Mould, "Auxilary Rules of Quantum Mechanics", quant-ph/0505231

\bibitem{RM2}R. A. Mould, ``Without the Born Rule", quant-ph/0507170

\bibitem{RM3}R. A. Mould, ``Quantum Brains: The oRules" \emph{AIP Conf. Proc.} \textbf{750}, 261 (2005); FPP3, V\"{a}xj\"{o}
University, Sweden, June 2004 

\bibitem{RM4}R. A. Mould, 		"Quantum Brain oRules" quant-ph/0406016

\bibitem{HE}H. Everett, ``Relative State Formulation of Quantum Mechanics", \emph{Rev. Mod. Phys.} \textbf{29} (1957) 

\bibitem{GRW}G. C. Ghirardi, A. Rimini, T. Weber, ``Unified dynamics for microscopic and macroscopic systems", \emph{Phys. Rev.
D} \textbf{34}, 470 (1986)

\bibitem{GPR}G. C. Ghirardi, P. Pearle, A. Rimini, ``Markov processes in Hilbert space and continuous spontaneous location of a
system of identical particles", \emph{Phys. Rev. A} \textbf{42}, 78 (1990)

\bibitem{RM5}R. A. Mould, ``Location \& the nRules", quant-ph/0509012

\bibitem{RM6}R. A. Mould, "The Cat nRules"; quant-ph/0410147

\bibitem{RM7}R. A. Mould, ``Hamiltonian Based nRules -- Time's Arrow", 

quant-ph/0507211

\end{thebibliography}
\end{document}